# HCQA: Hybrid Classical-Quantum Agent for Generating Optimal Quantum Sensor Circuits

Ahmad Alomari and Sathish A. P. Kumar
Department of Computer Science, Cleveland State University, Cleveland, OH USA, 44115
a.alomari@vikes.csuohio.edu; s.kumar13@csuohio.edu

*Abstract*—This study proposes an HCQA for designing optimal Quantum Sensor Circuits (QSCs) to address complex quantum physics problems. The HCQA integrates computational intelligence techniques by leveraging a Deep Q-Network (DQN) for learning and policy optimization, enhanced by a quantum-based action selection mechanism based on the Q-values. A quantum circuit encodes the agent current state using $R_y$ gates, and then creates a superposition of possible actions. Measurement of the circuit results in probabilistic action outcomes, allowing the agent to generate optimal QSCs by selecting sequences of gates that maximize the Quantum Fisher Information (QFI) while minimizing the number of gates. This computational intelligence-driven HCQA enables the automated generation of entangled quantum states, specifically the squeezed states, with high QFI sensitivity for quantum state estimation and control. Evaluation of the HCQA on a QSC that consists of two qubits and a sequence of $R_x$, $R_y$, and $S$ gates demonstrates its efficiency in generating optimal QSCs with a QFI of 1. This work highlights the synergy between AI-driven learning and quantum computation, illustrating how intelligent agents can autonomously discover optimal quantum circuit designs for enhanced sensing and estimation tasks.

*Impact Statement*—The HCQA introduces a hybrid AI-quantum framework for generating optimal QSCs, contributing to foundational advances in quantum metrology and intelligent quantum control. By integrating a DQN with quantum-based action selection, the HCQA learns to construct quantum circuits that achieve high QFI with reduced gate complexity. This approach demonstrates how reinforcement learning can guide quantum circuit synthesis in a goal-directed, data-efficient manner. While this work is demonstrated on a simplified two-qubit, noise-free simulation, it provides a proof of concept for how intelligent agents can autonomously learn and optimize QSCs. Technologically, this contributes to the growing field of Quantum Reinforcement Learning (QRL) and supports future exploration of scalable, noise-resilient extensions. We acknowledge that real-world deployment such as medical diagnostics, environmental monitoring, or quantum navigation—requires substantial advances in hardware and algorithmic scalability. As such, our claims are grounded in the potential of this work to serve as a steppingstone toward more complex, application-driven quantum systems. Future research will focus on generalizing HCQA to higher-qubit regimes and incorporating noise models to better align with experimental and industrial settings.

*Keywords*—*HCQA, DQN, QRL, QSC, QFI, Quantum Action Selection.*

## I. INTRODUCTION

Reinforcement Learning (RL) is a machine learning approach that allows autonomous intelligent agents to learn by directly interacting with an environment. These agents are rewarded for performing actions, and their goal is to find an optimal policy to maximize these rewards, which results in solving the task of the environment [1, 4, 31]. As artificial intelligence progresses, there is a greater need for algorithms that can learn rapidly and effectively, and speedups are more welcome than ever.

Quantum control refers to the use of classical or quantum RL agents to automatically design or optimize quantum circuits to address optimization tasks. The optimization objectives for the agent may include minimizing the number of gates, optimizing quantum states or entanglement, improving gate fidelity, and achieving other goals [1-14]. The agent uses optimization techniques to explore the vast design space of possible quantum circuits and selects designs that best meet the specified optimization objectives.

One set of metrologically useful states are squeezed states which give mild performance gains through pairwise entanglement generation [15]. In this case, a more complex sensor circuit is necessary to generate the quantum states that we need. The task therefore is to generate the specific kind of entanglement that will lead to an optimal quantum advantage for parameter estimation, for example, for the precision measurement of a phase shift. This phase shift could be generated by an inertial rotation, a magnetic field, or a variety of other possible terms in the system Hamiltonian of interest. The resulting technology will allow for a better understanding of the physical world with a breadth of applications that bridge many fields of science. We encapsulate this problem in a general conceptual framework referred to as a QSC [16, 17, 18].

A QSC executes a generalized Ramsey measurement on an array of qubits with encoding and decoding sequences represented by a chain of elementary gate operations, and the quantum design task is to select the optimal sequences [19]. The goal is to reveal the absolute quantum limit for measurement sensitivity when the circuit is taken as a whole. The design of such a QSC is difficult when the circuit is deep (meaning the cascade of many consecutive elementary gate operations) due to the large number of possible permutations of gates and measurements that should amplify the correct amplitudes for a sensitive signal while mitigating the adverse effects of noise and decoherence. While there are a variety of alternate approaches in optimal control theory, they all require a cycle of measurement and feedback, or open loop control, where exquisite modeling of the system is essential. This can lead to the degradation of the design utility when non-modelled perturbations are present [20]. These may include such imperfections as unitary errors due to gate inaccuracies, decoherence, dissipation, noise on control fields, and erasure errors of qubits.

In our proposed work, we developed an HCQA for quantum control tasks by integrating computational intelligence techniques. The HCQA leverages a DQN for learning and policy optimization, combined with a quantum-based action selection mechanism to generate optimal QSCs. By intelligently exploring action sequences, the HCQA efficiently produces entangled quantum states, specifically squeezed states, achieving high QFI values for enhanced quantum sensing and estimation. The QFI is an essential quantum mechanical measure of precision and sensitivity within quantum parameter estimation [21, 22, 23, 24]. The DQN is a multi-layered neural network that generates a vector of action values $Q(s,.;\theta)$ concerning a specific state $s$ [25]. The DQN consists of two networks: a regular Q-network $\theta$ and a target network $\theta'$. Here $\theta$ is used to select the action in each learning step, and $\theta'$ is used to evaluate the future optimal return.

The HCQA generates optimal QSCs by employing a quantum action selection method to select actions that maximize the QFI while minimizing the number of gates. By encoding the state information generated from the highest Q-value as rotation angles, the HCQA uses quantum superpositions and measurements to select the optimal quantum gate configuration for maximizing the QFI. High QFI indicates that the circuit is more sensitive to parameter changes and therefore more informative or useful for quantum state estimation or other quantum control tasks. Few gates mean that the circuit is not complex and can be implemented in quantum computers. The motivation of the proposed HCQA lies in addressing the limitations of classical DQN and quantum agents. Classical DQNs struggle with efficient exploration of large state spaces due to deterministic action selection, while quantum agents often face scalability challenges and slower convergence [1-14]. The proposed HCQA addresses these issues by integrating computational intelligence techniques, utilizing DQN for policy optimization and a quantum circuit for probabilistic action selection. This HCQA maximizes the QFI, while minimizing gate complexity, making it very efficient for generating optimal QSCs and enhancing quantum metrology and control tasks.

The remainder of the paper is organized as follows. Section II explains the current state of the art in the field of QRL. Section III describes the methodology for the proposed HCQA approach. Section IV describes the experimental results of the HCQA for quantum control tasks. Finally, section V concludes the paper.

## II. RELATED WORK

Existing QRL approaches can be classified into two types: quantum agents that learn in classical environments and scenarios where the agent and environment are fully quantum.

### A. Quantum Agents in Classical Environments

Examples of the first type are found in [1, 2, 4, 6, 8, 10, 14]. Heimann et al. developed a Deep Quantum Reinforcement Learning (DQRL) for training a wheeled robot to navigate through complex environments [1]. The wheeled robot is a Double Deep Q-Network (DDQN) that interacts with an environment represented using Variational Quantum Circuits (VQCs). The authors used the data-reuploading method to transform the classical features into the VQCs. The robot was tested using three different scenarios of the Turtlebot 2 environment, such that the complexity of the environment increased in each scenario. The results of the proposed DQRL show that quantum machine learning can be applied for autonomous robotic enhancements.

Skolik et al. developed a Variational Quantum Algorithm based on Deep Q-Learning (VQA-DQL) for enhancing the training process of Parametrized Quantum Circuits (PQCs) that can be used to solve discrete and continuous state space RL tasks [2]. The authors tested the proposed QRL approach using the frozen lake and cartpole environments, and the results show that QRL can outperform classical RL in terms of generating q-values for better learning performance agents.

Samuel et al. developed a hybrid quantum-classical approach that consists of quantum circuits with tunable parameters to enhance the performance of Noisy Intermediate Scale Quantum (NISQ) devices [6]. The proposed approach consists of VQCs that represent classical DRL algorithms (e.g., experience replay and target network). The circuits represent a Quantum Neural Network (QNN) that is used to estimate the Q-value function, which is used to improve the decision-making and policy selection of NISQ systems by reducing energy costs.

Yun et al. presented a Centralized Training and Decentralized Execution Quantum Multi-Agent Reinforcement Learning framework (CTDE-QMARL), which addresses challenges related to NISQ and non-stationary properties [8]. The proposed framework incorporates innovative VQCs and demonstrates its effectiveness in a single-hop environment, improving the performance of rewards received by agents compared to classical frameworks.

Chen et al. introduced two deep quantum reinforcement learning frameworks: one utilizes amplitude encoding for the CartPole problem, while the other employs a hybrid Tensor-Network VQC (TN-VQC) architecture to handle high-dimensional inputs of the MiniGrid [10]. The results show the advantage of parameter saving with amplitude encoding and the potential for quantum reinforcement learning on NISQ devices through efficient input dimension compression.

Sequeira et al. presented a low depth policy for a reinforcement learning agent in a VQC [14]. The study demonstrates an efficient approximation of the policy gradient with logarithmic sample complexity relative to the number of parameters. Empirical results confirm the comparable performance of the proposed VQC policy gradient to classical neural networks in benchmarking environments and quantum control, utilizing few parameters, while also investigating the barren plateau phenomenon in quantum policy gradients through analysis of the fisher information matrix.

Dong et al. developed a QRL approach based on quantum superposition and parallelism for autonomous state value updating of agents [4]. The proposed QRL technique represents the action as a quantum superposition state, such that the eigenstate of the action is obtained by randomly observing the superposition state according to the collapse postulate of quantum measurement. The eigen action (eigen state) probability is determined by the probability amplitude and parallelly updated according to rewards. The proposed QRL provides a balance between exploration and exploitation and can speed up the learning process through quantum parallelism.

### B. Quantum Agents in Quantum Environments

Examples of quantum agents learning in quantum environments are found in [3, 5, 7, 9, 11, 12, 13]. Alomari & Kumar proposed a Quantum Reinforcement Agent (QRA) to design optimal QSCs for complex quantum physics problems [5]. The QRA selects gate sequences that maximize QFI, while minimizing gate numbers, enabling the generation of entangled states like squeezed states. Their evaluation using two qubits and $R_x$, $R_y$, and $S$ gates achieved a QFI of 1, demonstrating the agent efficiency and potential in solving quantum physics problems.

Alomari & Kumar presented a Reinforcement Learning Autonomous Quantum Agent (ReLAQA) that integrates a Grover Autonomous Quantum Agent (GAQA) with a Quantum TicTacToe (QTTT) game to outperform classical methods in action selection using quantum techniques [7]. The ReLAQA demonstrated faster and more efficient performance than classical techniques using fewer computational resources. This work paves the way for future developments in quantum circuits for reinforcement learning robotics and metrological enhancements in NISQ devices.

Wu et al. implemented a Quantum Deep Deterministic Policy Gradient (QDDPG) algorithm for efficient resolution of classical and quantum sequential decision problems [9]. The proposed QDDPG enables one-shot optimization for generating control sequences to achieve arbitrary target states, surpassing the need for optimization per target state as required by standard quantum control methods. Additionally, the algorithm facilitates the physical reconstruction of unidentified quantum states.

Wiedemann et al. proposed a Quantum Policy Evaluation (QPE) method that combines amplitude estimation and Grover search for solving quantum reinforcement learning tasks, achieving quadratically greater efficiency compared to classical Monte Carlo methods [11]. Using QPE, the authors developed a Quantum Policy Iteration (QPI) approach that iteratively improves policies using the Grover search. The authors provide implementation and simulation for a two-armed bandit markov decision process to showcase the effectiveness of the proposed approach.

Borah et al. developed a Deep Reinforcement Learning Artificial Neural Agent (DRLANA) to control highly nonlinear quantum systems toward the ground state [12]. By incorporating weak continuous measurements into the proposed DRLANA, successfully learns counterintuitive strategies and achieves high fidelity. This approach demonstrates effective control techniques for nonlinear Hamiltonians, surpassing traditional methods.

Sivak et al. proposed a Model-Free Circuit-based Reinforcement Learning (MFCRL) approach for training an agent on quantum control tasks, addressing the issue of model bias [13]. The agent learns the parameters of a control PQC through trial-and-error interaction with the quantum system, utilizing measurement outcomes as the sole source of information. The proposed approach enables rewarding the agent using experimentally available observables, facilitating the preparation of nonclassical states, and executing logical gates on encoded qubits.

Dong et al. proposed the Quantum-inspired Reinforcement Learning (QiRL) algorithm for autonomous mobile robot navigation control [3]. The proposed technique uses a probabilistic action selection technique and a reinforcement policy, which are inspired, respectively, by the quantum measurement collapse phenomenon and amplitude amplification.

TABLE I
QRL APPROACHES

| Author | Approach | Environment Type | Advantages | Disadvantages | Metrics |
|---|---|---|---|---|---|
| Heimann et al. [1] | DQRL | Classical environment | Enhances the learning process of autonomous robotics applications | Qubit decoherence. The trainable parameters decrease the learning performance | Training steps, the number of trainable parameters, and average reward |
| Skolik et al. [2] | VQADQL | Classical environment | Provide an efficient decision policy | Qubit decoherence. Not a fully quantum agent. High complexity | Error median and Training steps |
| Samuel et al. [6] | RLVQCs | Classical environment | Enhances the learning process of VQCs | The implementation of the agent is not fully quantum. Classical computational errors in estimating the hyperparameters | Fidelity and accuracy |
| Yun et al. [8] | CTDE- | Classical environment | Limited applicability | Improves the performan | Average reward |

| Author | Approach | Environment Type | Advantages | Disadvantages | Metrics |
|---|---|---|---|---|---|
| | QMARL | | to complex domains | ce of rewards | |
| Chen et al. [10] | TN-VQC | Classical environment | Efficient data dimension compression | High complexity | Average reward |
| Wiedemann et al. [11] | QPE | Quantum environment | Efficient learning | Limited scalability for complex environments | Average reward |
| Sequeira et al. [14] | VQC | Classical environment | Efficient learning | Does not provide a balance between exploration and exploitation | Average reward |
| Dong et al. [4] | QRL | Classical environment | Provides a balance between exploration and exploitation | Not a fully quantum agent. Deals with small state space, which makes the learning process insufficient | Training steps |
| Alomari & Kumar [5] | QRA | Quantum environment | Provides a balance between exploration and exploitation | Not a fully quantum agent | QFI |
| Alomari & Kumar [7] | ReLAQA | Quantum environment | Provides a balance between exploration and exploitation | There is no phase estimation to generate efficient search space | Observed states |
| Wu et al. [9] | QDDPG | Quantum environment | Efficient one-shot optimization | Limited scalability for large-scale problems | Average reward |
| Borah et al. [12] | DRLANA | Quantum environment | Enhances quantum state control | High computational cost | Fidelity |
| Sivak et al. [13] | MFCRL | Quantum environment | Eliminates model bias | High complexity | Fidelity |
| Dong et al. [3] | QiRL | Quantum environment | Efficient learning performance | The representation of the state space is not accurate | Accuracy and learning rate |

Table I shows the surveyed QRL approaches and their limitations. In the proposed work, we developed an HCQA for generating optimal QSCs capable of overcoming the limitations discussed in Table I and solving complex quantum physics problems. Our proposed HCQA differs from the methods in [1-14] as it is not an enhanced quantum version of classical techniques; instead, it combines classical and quantum techniques to maximize the QFI of the generated QSCs while minimizing the number of gates.

## III. METHODOLOGY

Our proposed QRL approach consists of the HCQA and the QSC environment represented. Fig. 1 illustrates the workflow of the proposed QRL approach.

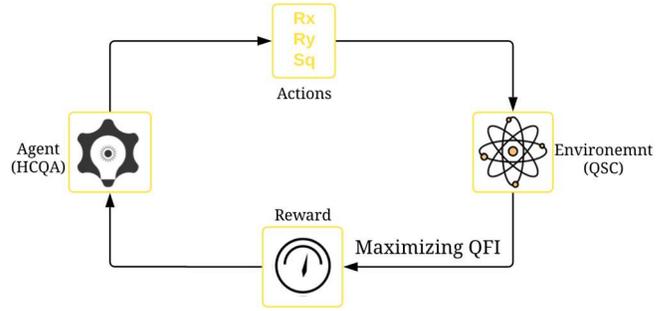

Figure 1. The Proposed QRL Workflow

The proposed QRL integrates quantum circuits with a DQN to generate optimal QSCs with high QFI and few gates. Each cycle starts with an environment reset, where Q-values guide the state-action decision-making. The Q-values, updated after each interaction, offer a measure of expected future rewards, informing the rotation angles within the quantum action selection circuit. By iteratively adjusting these angles based on Q-values, the HCQA selects actions that maximize the QFI while refining circuit efficiency. For action selection, the agent encodes the Q-values as rotation gates into a quantum circuit, then applies Hadamard ($H$) gates to create a superposition of actions, and measures the circuit to determine action probabilities. The agent selects the action with the highest probability, which can be $R_x$, $R_y$, and $S$ gates. This enhances exploration across multiple actions. Once an action is selected, it is applied to the environment, altering the quantum state. The environment then calculates the QFI of the selected action, which serves as the reward signal. The agent utilizes a DQN for policy learning, where Q-values are updated based on a classical Q-learning approach using a regular Q-network $\theta$ and a target network $\theta'$ [25]. The network is trained to predict the Q-values of state-action pairs, aiding the agent in refining its policy over time. This HCQA efficiently combines quantum circuits for action selection with a classical DQN for policy enhancement, illustrating its potential for generating optimal QSCs.

### A. The QSC Environment

To illustrate the efficient performance of the proposed QRA, we consider a QSC that consists of two qubits and $R_x$, $R_y$, and $S$ gates. Fig. 2 shows the structure of the QSC. This QSC has an optimal solution that we intend the QRA to generate, namely it should produce the N00N state since that maximizes the QFI [26].

First, we initialize the qubits to the state $|0\rangle$. Then, we apply a generalized Ramsey sequence to measure the QFI, which represents the accumulation of a relative phase between the two qubits collective dipole and a stable local oscillator that synchronizes the timing. This phase is associated with the generator of the Ramsey sequence such as the Pauli-Z operator ($Z$) [27, 28]. The accumulation consists of a sequence of $R_x$, $R_y$, and $S$ gates that the QRA optimally generates to maximize the QFI. The QFI and $S$ are given in equations 1 and 2 as follows.

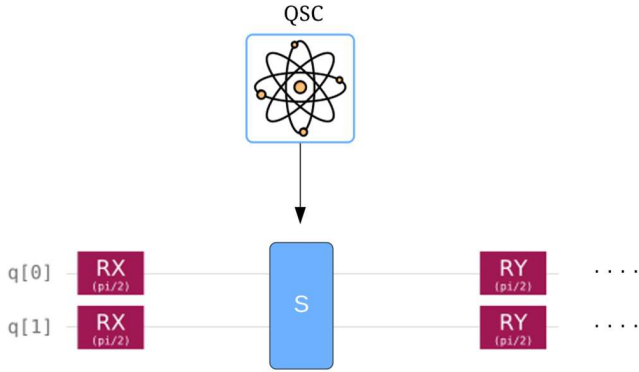

Figure 2. The Proposed QSC

$$QFI = \frac{4(\langle\psi|Z^2|\psi\rangle - \langle\psi|Z|\psi\rangle^2)}{n} \quad (1)$$

$n$ is the number of the qubits in the QSC, which is 2. $|\psi\rangle$ is the quantum state that is manipulated using the quantum gates that are generated by the QRA. $Z$ is the generator related to the rotation angle $\theta$. $Z$ represents the projection of the angular momentum of a quantum state along the z-axis by $\theta = \frac{\pi}{2}$. It is used to measure the z-component of the quantum state angular momentum.

In this work, we use the Quantum Fisher Information (QFI) as the primary evaluation metric, with a target value of 1 indicating an optimal quantum state configuration (QSC) characterized by high entanglement. A QFI value of 1 reflects the theoretical upper bound in our normalized setting, serving as a clear indicator of both quantum-enhanced sensitivity and the successful construction of maximally entangled states. We normalize QFI values to the [0, 1] range to ensure interpretability and to maintain a consistent and bounded reward structure throughout the learning process. This normalization simplifies convergence analysis and enables meaningful comparisons across episodes. A normalized QFI of 1 corresponds to the theoretical maximum (e.g., QFI = 4 in our two-qubit system) and thus reflects the agent's success in generating maximally entangled, information-rich quantum states. As such, convergence to QFI = 1 confirms that the learned policy reliably synthesizes optimal quantum sensor circuits within our defined framework.

$$S = \exp(-i\theta Z^2) \quad (2)$$

Due to the exponential scaling of the Hilbert space, a reduced representation of the QSC is necessary to represent the state in the QRL, and for this we use a discretized version of the Husimi-Q representation on the two qubits [29, 30]. The Husimi-Q function is a quasi-probability distribution in quantum mechanics that provides a phase-space representation of a quantum state. In our QSC, the discretized Husimi-Q function is calculated for a quantum state $|\psi\rangle$.

$$Q(\theta, \phi) = |\langle\alpha(\theta, \phi)|\psi\rangle|^2 \quad (3)$$

Where $|\alpha(\theta, \phi)\rangle$ is the quantum state parameterized by angles $\theta$ and $\phi$ in the Bloch sphere, and $|\psi\rangle$ is the quantum state. Upon completion, $Q(\theta, \phi)$ returns the updated quantum state by applying the gate that corresponds to the action that was passed when the $Q(\theta, \phi)$ was called. Then, we calculate the QFI, which determines the ability to resolve the accumulated relative phase $\phi$. It is the aim of the circuit overall to measure this phase with quantum limited sensitivity.

### B. The Proposed HCQA

The HCQA is designed to generate optimal QSCs with high QFI and few gates. It combines a DQN with a quantum action selection technique to enhance the learning process. Algorithm I details the workflow of the HCQA, demonstrating its iterative improvement and convergence toward actions that maximize the QFI.

To ensure reproducibility and provide transparency into our experimental setup, we detail the architecture and training parameters of the DQN agent used in our hybrid framework. The DQN comprises three fully connected layers with 128, 64, and 32 neurons, respectively, each followed by a ReLU activation function. The output layer produces Q-values corresponding to the discrete quantum gate actions. We use a learning rate of 0.001 and a discount factor $\gamma$ of 0.99. The agent follows an epsilon-greedy exploration strategy, where $\varepsilon$ is annealed linearly from 1.0 to 0.01 over 2000 episodes. Training is conducted using a mini-batch size of 64, and experiences are sampled from a replay buffer of size 10,000. The policy network is updated every 10 environment steps, while the target network is synchronized every 100 steps. These implementation details are chosen based on standard reinforcement learning practices and empirically validated through preliminary tuning.

ALGORITHM I
THE PSEUDOCODE OF THE HCQA

Initialize Environment and DQN: highest
  - Set parameters: $n = 2$, actions ($R_x$, $R_y$, $S$), $T$
  - Initialize DQN with input = Husimi-Q state, output = Q-value per action, learning rate $\alpha$, discount factor $\gamma$, reward $r$

Quantum_action_selection(state):
  - $\theta = Highest\ Q(a)$
  - Normalize $\theta$: $\theta = \frac{Highest\ Q(a)}{\sum_1^i Q(a_i)} \times \pi$
  - Apply $R_y(\theta)$ to both qubits and $H$ for superposition
  - Measure qubits in a computational basis
  - Convert the results to probabilities $P$, select the action with the highest $P$: $action = argmax(P(a))$

Training:
  - For each episode:

- Rest the state and episode actions
- Set done to false:
- While not done: $QFI > T$ or max steps reached
  - $quantum\_action\_selection(state)$
  - Apply Action:
    - Apply the gate to the QSC
  - Compute $QFI$: $QFI = \frac{4(\langle\psi|Z^2|\psi\rangle - \langle\psi|Z|\psi\rangle^2)}{n}$
- Update Q-values:
  - Compute target: $target = Q(s,a) + \alpha(r + \gamma * max(Q(s',a')))$
  - Update DQN weights using loss $(target - Q(s,a))$
- Accumulate $QFI$ and update state: $s \leftarrow s'$

Plot and Print: Plot the $QFI$ for all the episodes and print the optimal QSC

Algorithm I illustrates the interaction between the HCQA and the QSC environment. The episode starts from the training loop and ends when the termination condition is met. At the start of each episode, the agent initializes its state by resetting the environment. This reset brings the QSC back to its initial configuration, represented by calculating the initial value of the QFI.

The agent utilizes a quantum action selection method, which enables it to efficiently explore actions based on probabilistic outcomes. This method is a quantum circuit that consists of two qubits with two $R_y$ gates, two $H$ gates, and two measurement gates. The circuit determines the probabilities of the actions based on measurements. The rotation angle $\theta$ of the $R_y$ gates are encoded from the highest Q-value, where each action in Fig. 1 has its own Q-value. Fig. 2 shows the circuit of the quantum action selection technique.

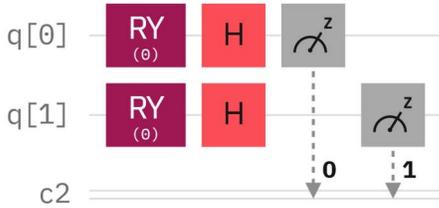

Figure 3. Quantum Action Selection Circuit

Fig. 3 illustrates the circuit of the quantum action selection technique. First, the circuit encodes the rotation angle $\theta$ of the $R_y$ gates based on the highest Q-value that is learned from the DQN, where each action in Fig. 1 has it is own Q-value. To ensure that $\theta$ has a range between $[0 - \pi]$, we normalize it is value using the following equation.

$$\theta = \frac{Highest\ Q(a)}{\sum_1^i Q(a_i)} \times \pi \quad (4)$$

We take the highest Q-value of an action $a$ obtained from the DQN and divide it over the sum of the Q-values of all actions $\sum_1^i Q(a_i)$, and then multiply it by $\pi$. Normalizing the rotation angle $\theta$ offers multiple advantages: it ensures compatibility with the $R_y$ gate range between $[0 - \pi]$, reflects the relative importance of actions by dividing the highest Q-value against the sum of all Q-values, and maintains stability by preventing unnecessary changes in angles during training [31]. This approach prioritizes high Q-value actions, aiding in effective decision-making, while facilitating better learning dynamics and efficient adaptation of the quantum action selection to evolving Q-values.

Next, the $R_y$ gates rotate the qubits around the y-axis by $\theta$, then the $H$ gates create a superposition of all the states, where each measurement outcome (00, 01, 10) corresponds to an action ($R_x, R_y, S$). Finally, we select the action that has the highest probability. For example, if 10 has the highest probability, then we select $S$ gate as the optimal action that will maximize the QFI.

Once the action is selected, it is applied to the QSC, transforming the quantum state through the $R_x, R_y, S$ gates. The new quantum state $\psi$ is then used to calculate the updated QFI using equation 1. Next, we update the Q-values of the actions using the following equation.

$$Q(s,a) = Q(s,a) + \alpha\left(r + \gamma\ max(Q(s',a') - Q(s,a))\right) \quad (5)$$

$Q(s,a)$ is the Q-value for the current state-action pair, $\alpha$ is the learning rate, $\gamma$ is the discount factor, $s'$ is the next state, and $r$ is the reward. The reward $r$ of the DQN at step i is represented by the following equation.

$$r_i^{DQN} = r_{i+1} + \gamma\ max_a Q(s',a;\theta_i') \quad (6)$$

Equation 6 indicates that the observed transitions of the experience replay environment are retained for some time and sampled uniformly from a memory bank to update the entire network [25]. This means that the correlation between the regular Q-network $\theta$ and the target network $\theta'$ dramatically improves the learning process through a weighted average optimization technique, which solves the over-optimism problem.

$T$ represents the threshold of the QSC environment, which is equal to 0.95. we terminate the episode when we generate a $QFI \geq T$ or we reach the maximum number of allowed actions in the episode, which is 10. $T$ indicates that the QSC generates high QFI values. A high QFI represents the successful generation of an optimal QSC that produces entangled quantum states. Otherwise, reaching the maximum number of allowed actions in an episode without achieving a QFI of 1 indicates that we are unable to generate an optimal QSC.

## IV. EXPERIMENTAL RESULTS

We implemented the proposed QRL approach, which consists of the HCQA and the QSC environment using local simulation. The simulation is done on an Intel Core I-7 Dell computer with 16 GB of RAM using Qiskit through Python. All experiments use the statevector simulator. Figs. 3, 4, 5, and 7 show the results of running the proposed HCQA, where the maximum number of allowed actions in an episode is 10. The results illustrate the efficient performance of the proposed HCQA in generating optimal QSCs. Furthermore, we compared the performance of the proposed HCQA with the QRA, GAQA, and classical DQN [5, 7, 25]. The QRA and the GAQA are quantum agents that we developed in our previous work, utilizing quantum techniques and the QFI to generate optimal QSCs. For more details about the QRA and the GAQA see [5, 7]. The results illustrate the efficient performance of the proposed HCQA in generating optimal QSCs compared to the other approaches.

## A. The Performance of the Proposed HCQA

To demonstrate the computational power of the proposed HCQA we ran it for three episodes. Figs. 4-6 show the experimental results that are generated using the statevector simulator, which assumes the absence of noise, and also eliminates sampling errors.

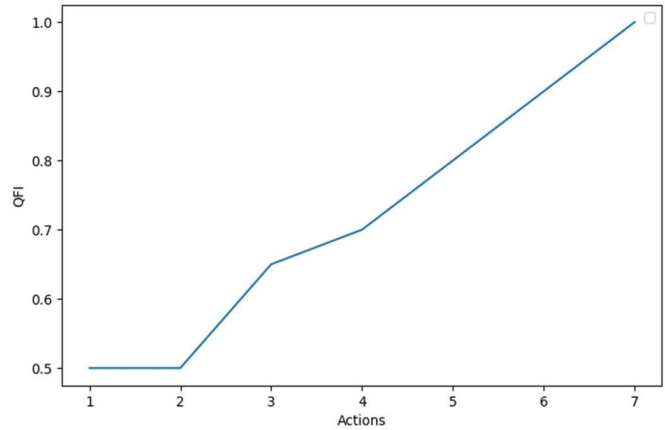

Figure 4. The Results of Episode 1

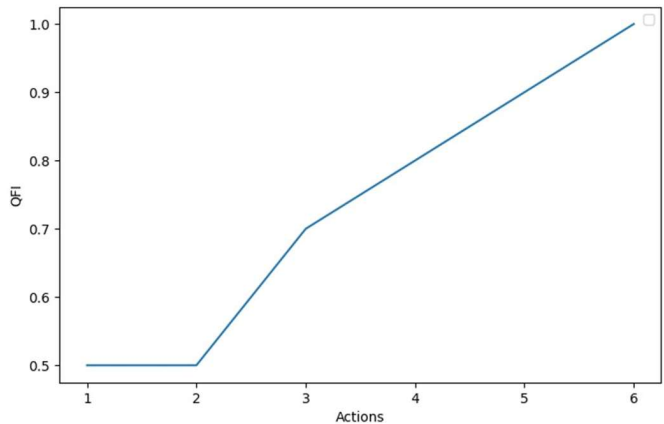

Figure 5. The Results of Episode 2

Figs. 4 and 5 show the results of episodes 1 and 2. In Fig. 4, the HCQA generated a QSC using 7 actions, with each quantum gate representing an action. In Fig. 5, the QSC is generated using 6 actions. It is important to note that the generated QSCs do not represent an optimal design for generating the N00N state because they are complex in terms of the number of gates [26]. However, both QSCs maximize the QFI, by generating a QFI of 1.

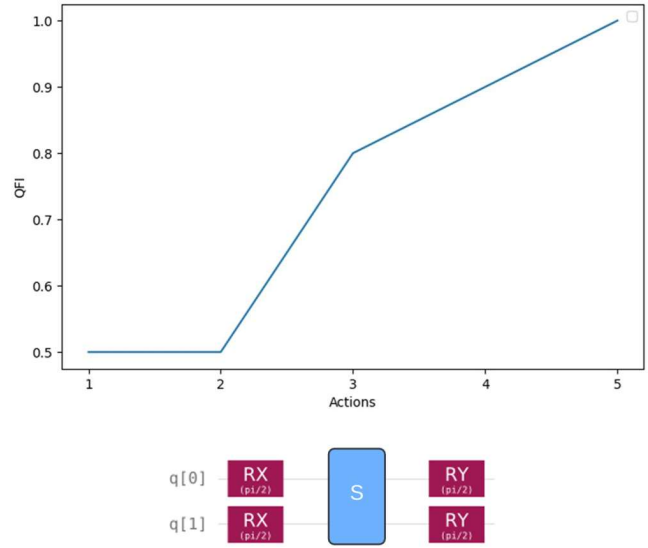

Figure 6. The Results of Episode 3

Fig. 6 shows the results of episode 3. The HCQA generated a QSC using 5 actions. The generated QSC represents an optimal design for generating the N00N state because less complex in terms of the number of gates and maximizes the QFI by generating a QFI of 1.

## B. The Results of Comparing the HCQA with the Other Approaches

We compared the performance of the HCQA with the QRA, GAQA, and classical DQN [5, 7, 25]. The GAQA is a quantum agent that utilizes Grover search and amplitude amplification. Fig. 7 shows the workflow of the GAQA.

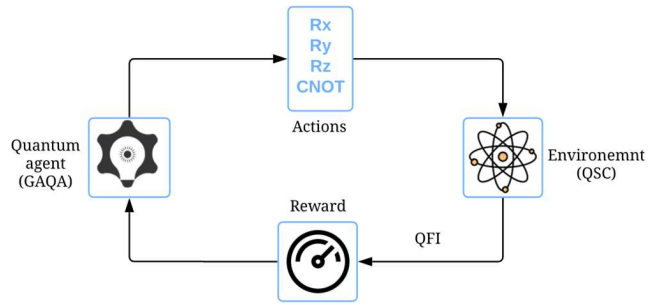

Figure 7. The Workflow of the GAQA

Fig. 7 shows the workflow of the GAQA. In our previous work, we used the GAQA to solve complex robotics applications, such as the QTTT environment [5]. The problem is to find the optimal quantum circuit that solves the QTTT. In this experiment, we modified the actions to those shown in Fig. 9. We want the GAQA to find the optimal QSC that is represented by qubits 0-2 in Fig. 8. The GAQA starts by selecting an action, which is a gate that can be $R_x$, $R_y$, $R_z$, and

$CNOT$. It then generates the QFI of the selected action using equation 1. Next, the GAQA utilizes amplitude amplification to increase the probability amplitude of the selected action. It then uses this probability and the generated QFI as a reward to select the next action that maximizes the QFI. This process continues until the GAQA generates a QFI of 1 or reaches the number of actions allowed in an episode which is 10. Modifying the GAQA is not a straightforward application since we had to change the actions to generate optimal QSCs [7]. We did not modify the QRA for this performance evaluation because it runs over the QSC as we describe in [5]. The results are shown in Fig. 8.

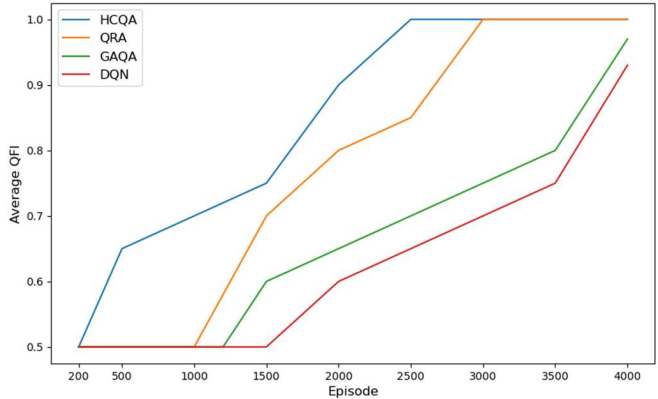

Figure 8. The Average QFI of the HCQA vs Other Approaches

Fig. 8 illustrates the generated average QFI of the HCQA compared to the QRA, GAQA, and classical DQN over 4000 episodes. The HCQA generated an average QFI of 1, outperforming the other approaches. All agents generated a QFI of 0.5 from 1 to 200, and then the HCQA started to generate higher QFI values from episodes 200 to 4000. The HCQA generated QFI values of 1 starting from episode 2500 to 4000, which means that it kept generating the optimal QSC in Fig. 6 during these episodes. The QRA has the second best performance as it started to generate a QFI of 1 from episode 3000 to 4000. The GAQA did not generate a QFI of 1 but it started to generate higher QFI values than the classical DQN from episode 1500 to 4000, reaching a QFI of 0.95. The classical DQN has the worst performance, generating the lowest QFI values from episodes 1200 to 4000 and reaching a QFI of 0.92. The results demonstrate that the HCQA has a better performance than the other approaches.

Moreover, we compared the performance of the proposed HCQA with the Grover Policy Agent (GPA), which is our latest quantum agent [32]. The proposed GPA consists of the Quantum Policy Evaluation (QPE) and the Quantum Polic Improvement (QPI). The idea of the QPE is to generate the search space, which consists of the evaluated policies. QPI then uses Grover search and amplitude amplification techniques to find an optimal policy that will be used to generate QSCs. The main difference between the proposed GPA and the GAQA is that the GPA uses phase estimation to generate the policies, which reduces the search space. Also, the GPA generates the QFI using our proposed subtraction operations in [32]. These enhancements are not implemented in the GAQA. For more details about the GPA see [32]. The results are shown in Fig. 9.

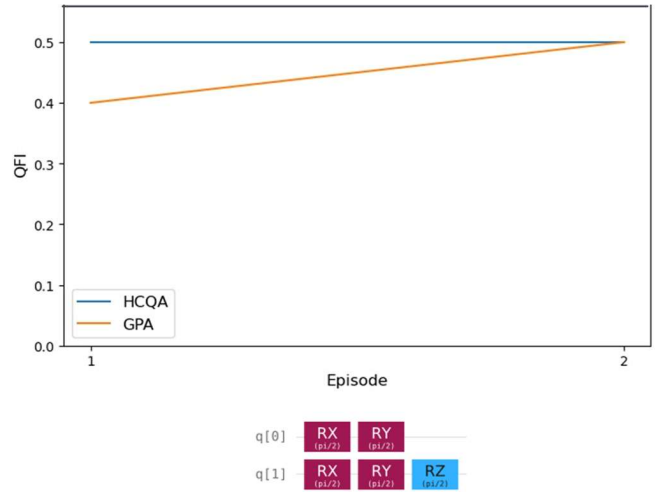

Figure 9. The Average QFI of the HCQA vs GPA

Fig. 9 shows the generated average QFI of the HCQA compared to the GPA over 2 episodes. The GPA is a pure quantum agent, meaning that it does not use classical reinforcement learning approaches. It operates using quantum techniques, with episodes running in parallel rather than sequentially, as in classical approaches. This significantly increases the complexity of the GPA. To control this complexity, we simplified the QSC in Fig. 2 by removing the squeezing gates $S$ and generating the QFI using our proposed subtraction operation. This allows us to run the GPA for two episodes, controlling its high complexity. For more details about the simplified QSC see [32]. To ensure a fair comparison, we ran the GAQA for two episodes, matching the GPA limitation of only two episodes. Since there are no squeezing $S$ gates, the maximum QFI of the QSC will be 0.5, resulting form $R_x$, $R_y$, and $R_z$ gates. We added the $R_z$ gate, which is a rotation around the z-axis by $\theta = \frac{\pi}{2}$ as a generator to make sure that the subtraction operation of the GPA generates the QFI as described in equation 1 [32]. The HCQA generated a QFI of 0.5 in both episodes, outperforming the GPA, which generated a QFI of 0.4 in the first episode and a QFI of 0.5 in the second episode. The QSC in Fig. 9 is the optimal QSC that is generated by the HCQA in the second episode. The results illustrate that the HCQA has a better performance than the GPA.

V. CONCLUSION

Quantum control involves the utilization of classical or quantum RL agents for designing and enhancing quantum circuits to address optimization challenges. The optimization objectives for the agent may include minimizing the number of gates, optimizing quantum states or entanglement, improving gate fidelity, and achieving other goals. In this proposed work, we have developed a HCQA approach for generating optimal QSCs capable of solving complex quantum physics problems by generating entangled quantum states, specifically the squeezed states. The QSC performs a generalized Ramsey measurement on qubits using sequences of quantum gates, and the task of the QRA is to select the optimal sequences. The aim is to generate the maximum QFI when the circuit is taken as a whole. The proposed HCQA utilizes a DQN for learning and policy optimization, enhanced by a quantum action selection technique based on the Q-values. A quantum circuit encodes the agent current state using $R_y$ gates, and then creates a superposition of possible actions. Measurement of the circuit allows the agent to generate optimal QSCs by selecting actions that maximize the QFI, while minimizing the number of gates. High QFI indicates that the circuit is more sensitive to parameter changes and therefore more informative or useful for quantum state estimation or other quantum control tasks. Few quantum gates mean that the circuit is not complex and can be implemented in quantum computers.

To evaluate the performance of the proposed HCQA, we considered a QSC that consists of two qubits and a sequence of $R_x$, $R_y$ and $S$ gates. This circuit has an optimal N00N state that maximizes the QFI, and the task of the HCQA is to generate this circuit while minimizing the number of gates. The results show the efficient performance of the proposed HCQA by generating optimal QSCs with a QFI of 1. The implementation details and simulations we conducted will illustrate how quantum agents can be utilized to solve quantum physics problems. For future work, we intend to test the proposed HCQA approach on different quantum physics problems, which require the design of complex QSCs

ACKNOWLEDGMENTS

This work was supported by the National Science Foundation Grant No. OMA 2231377.